\documentclass[a4paper,11pt]{article}

\usepackage{amsmath}
\usepackage{amssymb}
\usepackage[dvips]{graphicx}
\usepackage{feynmf}
\renewcommand{\title}[1]{\null\vspace{10mm}\noindent
                         {\Large{\bf #1}}\vspace{8mm}}
\newcommand{\authors}[1]{\noindent{\large #1}\vspace{3mm}}
\newcommand{\address}[1]{{\center{\noindent\small\itshape #1\vspace{0mm}}}}

\allowdisplaybreaks[3]

\begin{document}

\newtheorem{OnShellPhasesOnly}{}

\begin{titlepage}

\begin{flushright}
hep-th/0306099 \\
CERN-TH/2003-126   

\end{flushright}

\begin{center}

\title{No UV/IR Mixing in Unitary Space-Time Non-Commutative Field Theory}

\authors{P.~Fischer$^1$, V.~Putz$^2$}

\address{$^1$Theory Division, CERN, CH-1211 Geneva 23, Switzerland}

\address{$^{2}$Max-Planck-Institut f\"ur 
Mathematik in den Naturwissenschaften, 
\\
Inselstra\ss{}e 22-26, D-04103 Leipzig, Germany}

\vspace{3cm}

\begin{abstract}
In this article we calculate several divergent amplitudes in $\phi^4$-theory on non-commutative space-time ($\Theta_{0i} \neq 0$)
in the framework of Interaction-Point Time-Ordered Perturbation Theory (IPTOPT),
continuing work done in hep-th/0209253.    
On the ground of these results we find corresponding Feynman rules that allow for a much easier diagrammatic 
calculation of amplitudes.
The most important feature of the present theory is the absence of the UV/IR mixing problem in all amplitudes 
calculated so far. Although we are not yet able to give a rigorous proof, we provide a strong argument for this
result to hold in general. 
Together with the Feynman rules we found, this opens promising vistas onto the systematic renormalization of 
non-commutative field theories. 
\end{abstract}

\end{center}

\footnotetext[1]{pfischer@mail.cern.ch}

\footnotetext[2]{putz@hep.itp.tuwien.ac.at, work supported by the ``Fonds zur
  F\"orderung der Wissenschaften'' (FWF) under contract P15015-TPH.}

\end{titlepage}

\section{Introduction}


In quantum field theories on non-commutative spaces,
we know of two major problems. The first one is the famous so-called 
``UV/IR mixing''. In using 
standard perturbative techniques a completely new type of non-renormalizable, 
infrared-like singularities occurs \cite{Minwalla:1999px}, 
\cite{Matusis:2000jf}. Attempts to cure this imponderability have 
been made, but no convincing solution has been found so far.

The second problem is the loss of unitarity on non-commutative spaces with
Minkowskian signature \cite{Bahns:2002vm}, \cite{Gomis:2000zz}. The first
resolutions made considerable, undesirable restrictions (e.g.~commutativity 
of time) and were thus not very satisfying. In
\cite{Doplicher:1994tu} and \cite{Rim:2002if} two proposals to cure this 
severe problem have been made. A similar approach was elaborated in 
\cite{Liao:2002xc}, \cite{Liao:2002pj}. There the Gell-Mann--Low formula 
for Green's functions:
\begin{eqnarray*}
&&G_n(x_1,...,x_k)\!:=\!\frac{i^n}{n!}\!\int\! d^4z_1...d^4z_n\big\langle 0|T
\phi(x_1)...\phi(x_k){\mathcal L}_I(z_1)...{\mathcal L}_I(z_n)
|0\big\rangle^{con},
\end{eqnarray*}
was used. 
The theory is quantized canonically in Minkowski space instead of employing the Euclidean path-integral (PI).
As shown in \cite{Bahns:2002vm}, \cite{Rim:2002if} and \cite{Liao:2002pj}, unitarity is recovered 
by choosing the Lagrangian as the starting point of this formulation of non-commutative field theories.
It must be stressed that for non-commutative time ($\Theta_{0i} \neq 0$) a theory different from the usual 
approach is considered and new results are to be expected.  

Technically this can easily be seen by noting that time ordering (TO) -- which contains $\Theta(x^0-y^0)$ -- 
stands \emph{in front} of the $*$-product, which, in the formulation $f(x)*g(x):= e^{i/2 \partial_{x,\mu} \Theta^{\mu \nu} 
\partial_{y,\nu}} f(x) g(y)|_{x=y}$, 
contains  an infinite number of time derivations for $\Theta_{0i} \neq 0$.
The consequences of this rather obvious fact were investigated in detail in \cite{Liao:2002xc}, \cite{Liao:2002pj} for
$\phi^3$ and $\phi^4$ in \cite{Noncom1} for $\phi^4$.

In this last work it was also shown that if one employs another definition of the $\star$-product,
\begin{eqnarray}
&&{\mathcal L}_I(z_l) = \frac g{4!}(\phi\star\phi\star\phi\star\phi)(z_l)
= \int \prod_{i=1}^3\Big(d^4s_i \frac{d^4l_i}{(2\pi)^4}e^{il_i s_i}\Big)\\
&& \times\phi(z_l-\frac{\tilde l_1}{2})\phi(z_l+s_1-\frac{\tilde l_2}{2})
\phi(z_l+s_1+s_2-\frac{\tilde l_3}{2})\phi(z_l+s_1+s_2+s_3),\nonumber
\end{eqnarray}
a physical interpretation of the ensuing techniques is possible 
(aside from making the calculations easier and more transparent). 
Non-commutativity can be seen explicitly to ``spread'' the interaction over space-time.
The time ordering only acts on the time-stamp of the interaction point (IP) $z^0_l$, but not on the new, 
smeared-out ``physical'' coordinates of the field operators.
Thus the four fields of the interaction point are not time-ordered
with respect to each other,
time ordering being realized between external and interaction points only. 
This fact gave reason to the notion
of \emph{Interaction-Point Time-Ordered Perturbation Theory} (IPTOPT) introduced in \cite{Noncom1}, to distinguish from 
a true causal time ordering.
The fields at the interaction point are not causally connected, and ``micro-'' (better ``nano-'') causality is
violated at the non-commutative vertex \cite{Noncom1}, \cite{Bahns:2003vb}.  

In \cite{Noncom1} this approach was developed into IPTOPT in analogy 
to pre-Feynmanian commutative perturbation theory \cite{Sterman93}. 
The techniques developed so far are still rather cumbersome (though examples of their 
applicability are given below) and true diagrammatics including the respective Feynman rules (FR) 
are the next step in the implementation of this program.

This is undertaken in the present work (see also \cite{Denk}), which already rewards us with a possible solution to the second 
great problem of non-commutative field theories, UV/IR mixing.   
\newline{}





To reach these goals we set out from previous work. 
In section \ref{Examples} we employ the non-commutative version of the time ordered expression for 
Green functions \cite{Sterman93},
eq.~(39) of \cite{Noncom1}, to obtain explicit results for the Fourier-transformed (FT), amputated on-shell 
two-point one-loop amplitude $\Gamma^{(2,1)}$ (tadpole, fig.~\ref{oneloop}), 
two-point two-loop amplitude $\Gamma^{(2,2)}$ (snowman, fig.~\ref{twoloop}),
and  four-point one-loop amplitude $\Gamma^{(4,1)}$ (fish, fig.~\ref{fourpoint}).   
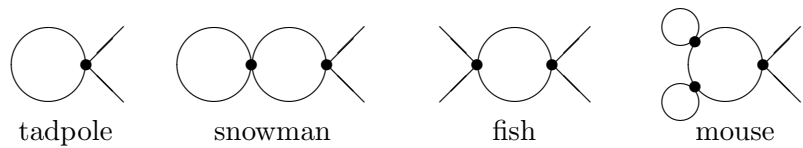
\begin{eqnarray*}
\begin{picture}(22,20)
\put(10,10){\circle{10}}
\put(14,9){\mbox{$\bullet$}}
\put(15,10){\line(1,1){5}}
\put(15,10){\line(1,-1){5}}
\put(6,0){tadpole}
\end{picture}
\begin{picture}(30,20)
\put(10,10){\circle{10}}
\put(14,9){\mbox{$\bullet$}}
\put(20,10){\circle{10}}
\put(24,9){\mbox{$\bullet$}}
\put(25,10){\line(1,1){5}}
\put(25,10){\line(1,-1){5}}
\put(10,0){snowman}
\end{picture}
\begin{picture}(28,20)
\put(15,10){\line(-1,1){5}}
\put(15,10){\line(-1,-1){5}}
\put(14,9){\mbox{$\bullet$}}
\put(20,10){\circle{10}}
\put(24,9){\mbox{$\bullet$}}
\put(25,10){\line(1,1){5}}
\put(25,10){\line(1,-1){5}}
\put(17,0){fish}
\end{picture}
\begin{picture}(30,20)
\put(14,15){\circle{5}}
\put(14,5){\circle{5}}
\put(15,6){\mbox{$\bullet$}}
\put(15,12){\mbox{$\bullet$}}
\put(20,10){\circle{10}}
\put(24,9){\mbox{$\bullet$}}
\put(25,10){\line(1,1){5}}
\put(25,10){\line(1,-1){5}}
\put(16,0){mouse}
\end{picture}
\end{eqnarray*}
In section \ref{Diagrammatics} we return to the result for the off-shell non-amputated Green function $G^{(2,1)}$ 
obtained in \cite{Noncom1}
by explicitly commuting out the free field operators. Retracing one step, we explicitly state the
full off-shell amplitude, the correction to the propagator at one loop. 

This result allows us to ``read off'' the TO propagator of our theory and the algorithm that allows us the 
construction of general diagrams.

The last missing item, the vertex, is easily obtained and completes the set of FR of IPTOPT.  

Section \ref{IPTOPTFR} is devoted to a demonstration of the correctness 
and applicability of our new FR by employing them in redoing the calculations of section \ref{Examples}. 
Of special interest to the issue of UV/IR mixing is a certain two-point three-loop amplitude $\Gamma^{(2,3)}$
(two tadpoles inserted into a third, 
the so-called \emph{mouse}-diagram of fig.~\ref{mouse}), where it generates new divergences. 
We calculate this expression in section \ref{MouseFR}.
 
The discussion of section \ref{Discussion} is mainly dedicated to what our results tell us about the UV/IR problem.
First we note that it does not appear anywhere in the determined amplitudes, especially not in $\Gamma^{(2,3)}$
which remains -- in contrast to the results normally obtained in non-commutative field theory -- void of new divergences.
This most interesting feature of the present theory encourages us to put forth a general argument for the absence of this 
notorious problem in IPTOPT, which we do in section \ref{NoUVIR}, at least in its usual form.
A short remark on the PT invariance of the obtained amplitudes is made in section \ref{PT}.

In the Outlook, section \ref{Outlook}, we give lines along which a rigorous proof (or disclaim) of the general absence of 
UV/IR mixing in IPTOPT may proceed. We also list the next steps in the program of IPTOPT, among which are of course
the attempt at a renormalization of this non-commutative field theory.

\section{Examples}   \label{Examples}

Now we want to look at some prominent diagrams with the help of
eq.~(39) of \cite{Noncom1}: 
\begin{align}  
&\Gamma(q_1^{\sigma_1},\dots,q_E^{\sigma_E}) \nonumber\\
&= \lim_{\varepsilon\to 0}\frac{g^V}{(4!)^V}\!
\int\! \prod_{i=1}^I \frac{d^3 k_i}{(2\pi)^3 2\omega_{k_i}}
\prod_{v=1}^{V-1} \frac{i
(2\pi)^3 \delta^3 \Big(\sum_{i=1}^I J_{vi} \vec{k}_i 
+\sum_{e=1}^E J_{ve}\sigma_e \vec{q}_e\Big)
}{
\sum_{v'\leq v} \!\Big( \sum_{i=1}^I J_{v'i} \omega_{k_i} 
+ \sum_{e=1}^E J_{v'e} \omega_{q_e}\Big)+i\varepsilon}
\nonumber
\\*
& \times 
\exp\Big(i \theta^{\mu\nu} \Big(
\sum_{i,j=1}^I I_{ij} k_{i,\mu}^+ k_{j,\nu}^+ 
+ \! \sum_{i=1}^I \sum_{e=1}^E I_{ie} 
\sigma_e k_{i,\mu}^+ q_{e,\nu}^{\sigma_e} 
+ \! \sum_{e,f=1}^E I_{ef} \sigma_e \sigma_f q_{e,\mu}^{\sigma_e}
q_{f,\nu}^{\sigma_f}\Big)\Big).
\label{GammaE}
\end{align}
The vertex that is missing in the product over $v$ is the latest one. Note
that this formula, because of the somewhat unusual definition of the S-matrix used in 
\cite{Noncom1}, has some extra factors $i$ with respect to the 
usual expression.

Here the internal (carrying the momenta $k$) 
and external (carrying the momenta $q$) lines are oriented forward in time
(note, however, that the external $momenta$ are always defined outgoing
of a vertex).
Then, the incidence matrices $J_{vi}, J_{ve}$ are equal to $-1$ if the line
leaves $v$ and to $+1$ if the line arrives at $v$. Similarly,
$\sigma_e=-1$ if the line $e$ leaves $x_e$ and $\sigma_e=+1$ if the
line $e$ arrives at $x_e$.  The matrices $I_{ij}, I_{ie}, I_{ef}$ are
the intersection matrices, which
describe the time configuration of the lines at a vertex. They will be defined below.

\subsection{Two-Point One-Loop Tadpole}

To see how the formula works, we first want to review
the on-shell one-loop correction to the two-point function. One typical 
contribution to this diagram is:
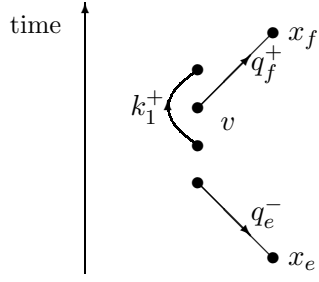
\begin{figure}[h]
\begin{center}
\begin{picture}(50,40)  
\put(19,14){\mbox{$\bullet$}}
\put(19,19){\mbox{$\bullet$}}
\put(19,24){\mbox{$\bullet$}}
\put(19,29){\mbox{$\bullet$}}
\put(29,4){\mbox{$\bullet$}}\put(32,4){\mbox{$x_e$}}
\put(29,34){\mbox{$\bullet$}}\put(32,34){\mbox{$x_f$}}
\put(20,15){\line(1,-1){10}}\put(27,10){\mbox{$q_e^-$}}
\put(20,25){\line(1,1){10}}\put(27,30){\mbox{$q_f^+$}}
\put(5,3){\vector(0,1){36}}
\put(16,25){\vector(0,1){1}}
\put(20,15){\vector(1,-1){7}}
\put(20,25){\vector(1,1){7}}
\put(-5,35){\mbox{\small time}}
\bezier{0}(20,20)(12,25)(20,30)\put(11,24){\mbox{$k_1^+$}}
\put(23,22){\mbox{$v$}}
\end{picture}
\caption{The contribution $(e,\bar1,f,1)$ \label{oneloop}}
\end{center}
\end{figure}
With eq.~(\ref{GammaE}) a general contribution reads
\begin{eqnarray}
\Gamma^{(2,1)} = 
\frac g{4!}\int \frac{d^3k}{(2\pi)^32\omega_k}\cdot1\cdot
\mathrm{exp}(i\theta^{\mu\nu}\phi_{\mu\nu}), \label{Gone}
\end{eqnarray}
where $\Phi_{\mu\nu}$ is the phase depending on the special configuration
of lines at the vertex. Since there is only one inner line, the $I_{ij}$ term 
in (\ref{GammaE}) is vanishing. 
For the $I_{ie}$ term we have to look at all possible 
configurations of lines at the 4-field vertex $v$. We have 
\begin{equation}
I_{ij}=\frac12\sum_{v}\tau^v_{ij}J_{vi}J_{vj},
\quad I_{ie} = \frac12\sum_{v}(\tau^v_{ie} - \tau^v_{ei})J_{vi}J_{ve},
\quad I_{ef} = \frac12\sum_{v}\tau^v_{ef}J_{ve}J_{vf}.
\end{equation}
The sum is over all vertices in a particular graph;
$\tau_{ie}^v=+1$ if the line $i$ is connected to an 
``earlier'' field $\phi$ in the vertex $v$ than the line $e$, otherwise
$\tau^v_{ij} = 0$. We have $\sigma_e = -1$, $J_{ve} =+1$, $\sigma_f = +1$, 
$J_{vf} = -1$. For the inner line
we have to distinguish between the one leaving (we denote this by $i = \bar1$) and
the one arriving ($i = 1$). Then $J_{v\bar 1} = -1$ and $J_{v1} = +1$. Note that the
inner line is by definition oriented forward in time and $k_1\equiv k_{\bar1}$.
We write the time-ordering configuration at the vertex as an array, the
contribution in figure \ref{oneloop} is labelled $(e,\bar1,f,1)$.
Then we find, for the $I_{ie}$ and the $I_{ef}$ terms:
\begin{eqnarray*}
&&\sum_{i=1,\bar1}k_i^+I_{ie}(-q^-_e) + \sum_{i=1,\bar1}k_i^+I_{if}(+q^+_f)
+ \sum_{e',f'=e,f}I_{e'f'}(\sigma_{e'}q_{e'}^{\sigma_{e'}})
(\sigma_{f'}q_{f'}^{\sigma_{f'}}) 
=\\
&&(e,f,\bar1,1): 0+0+\frac12 q_e^-q_f^+
\hspace{2.7cm}(f,e,\bar1,1): 0+0+\frac12 q_f^+q_e^- \\
&&(\bar1,1,e,f):0+0 +\frac12 q_e^-q_f^+
\hspace{2.7cm}(\bar1,1,f,e): 0+0+\frac12 q_f^+q_e^-  \\
&&(e,\bar1,1,f): 0+0+\frac12 q_e^-q_f^+
\hspace{2.7cm}(f,\bar1,1,e): 0+0+\frac12 q_f^+q_e^-  \\
&&(\bar1,e,f,1): -k_1^+(-q^-_e) + k_1^+q^+_f+\frac12 q_e^-q_f^+\\&& 
\hspace{5.1cm}(\bar1,f,e,1): -k_1^+(-q^-_e) + k_1^+q^+_f +\frac12 q_f^+q_e^- \\
&&(e,\bar1,f,1): +k_1^+q^+_f+\frac12 q_e^-q_f^+ 
\hspace{2.3cm}(\bar1,f,1,e): +k_1^+q^+_f+\frac12 q_f^+q_e^-\\ 
&&(\bar1,e,1,f): -k_1^+(-q^-_e)+\frac12 q_e^-q_f^+\hspace{1.1cm}
(f,\bar1,e,1): -k_1^+(-q^-_e) +\frac12 q_f^+q_e^-.
\end{eqnarray*}
Thus for the sum over all possible phase factors, we obtain
\begin{eqnarray}
&&\sum_{\phi} \mathrm{exp}(i\theta^{\mu\nu}\phi_{\mu\nu}) = 
2 \cos \big(\frac12\theta^{\mu\nu}q_{e,\mu}^-q_{f,\nu}^+ \big)\\&&
\times \Big(3 \mathrm{e}^0 + \mathrm{e}^{i\theta^{\mu\nu}k_{1,\mu}^+q_{e,\nu}^-}
+\mathrm{e}^{i\theta^{\mu\nu}k_{1,\mu}^+q_{f,\nu}^+}
+\mathrm{e}^{i\theta^{\mu\nu}(k_{1,\mu}^+q_{e,\nu}^-+
k_{1,\mu}^+q_{f,\nu}^+)}\Big)  . \nonumber\label{tad}
\end{eqnarray}
Inserting this into (\ref{Gone}) and with 
$q_f^+ = -q_e^-$ we find for the total $\Gamma$
\begin{equation}
\Gamma_{tot}^{(2,1)} = 
\frac g{12}\int \frac{d^3k}{(2\pi)^32\omega_k}\big(4+2
\cos(\theta^{\mu\nu}k^+_\mu q_{f,\nu}^+)\big).
\end{equation}
This result agrees with eq.~(25) of \cite{Noncom1}, where the same amplitude was 
obtained by explicitly commuting out the free field operators.

\subsection{Two-Loop Snowman} \label{Snowmanexact}

For the two-loop snowman, in addition to the inner configuration of
the lines at the vertices, we have to respect the two possibilities of
time ordering of the vertices:

 \begin{figure}[h] 
\begin{picture}(55,60)  
\put(19,34){\mbox{$\bullet$}}
\put(19,39){\mbox{$\bullet$}}
\put(19,44){\mbox{$\bullet$}}
\put(19,49){\mbox{$\bullet$}}
\put(39,10){\mbox{$\bullet$}}
\put(39,14){\mbox{$\bullet$}}
\put(39,18){\mbox{$\bullet$}}
\put(39,22){\mbox{$\bullet$}}
\put(40,11){\line(1,-1){10}}\put(40,11){\vector(1,-1){5}}
\put(40,23){\line(1,3){10}}\put(40,23){\vector(1,3){5}}
\put(49,0){\mbox{$\bullet$}}\put(52,0){\mbox{$x_e$}}
\put(49,52){\mbox{$\bullet$}}\put(52,52){\mbox{$x_f$}}
\put(20,35){\line(1,-1){20}}\put(29,18){\mbox{$k_2^+$}}
\put(20,45){\line(3,-4){20}}\put(30,34){\mbox{$k_3^+$}}
\put(8,3){\vector(0,1){50}}
\put(16,45){\vector(0,1){1}}
\put(30,25){\vector(-1,1){1}}\put(45,7){\mbox{$q_e^-$}}
\put(30,32){\vector(-3,4){1}}\put(45,33){\mbox{$q_f^+$}}
\put(-2,45){\mbox{\small time}}
\bezier{0}(20,40)(12,45)(20,50)\put(11,44){\mbox{$k_1^+$}}
\put(24,42){\mbox{$v$}}\put(43,16){\mbox{$w$}}
\end{picture}
\qquad
\begin{picture}(55,60)  
\put(19,16){\mbox{$\bullet$}}
\put(19,20){\mbox{$\bullet$}}
\put(19,24){\mbox{$\bullet$}}
\put(19,28){\mbox{$\bullet$}}
\put(39,33){\mbox{$\bullet$}}
\put(39,36){\mbox{$\bullet$}}
\put(39,39){\mbox{$\bullet$}}
\put(39,42){\mbox{$\bullet$}}
\put(20,17){\line(1,1){20}}\put(40,43){\vector(1,1){5}}
\put(20,25){\line(4,3){20}}\put(40,34){\vector(1,-3){5}}
\put(49,3){\mbox{$\bullet$}}\put(52,3){\mbox{$x_e$}}
\put(49,52){\mbox{$\bullet$}}\put(52,52){\mbox{$x_f$}}
\put(40,43){\line(1,1){10}}\put(30,24){\mbox{$k_2^+$}}
\put(40,34){\line(1,-3){10}}\put(29,36){\mbox{$k_3^+$}}
\put(8,3){\vector(0,1){50}}
\put(16,25){\vector(0,1){1}}
\put(30,27){\vector(1,1){1}}\put(47,17){\mbox{$q_e^-$}}
\put(31,33){\vector(4,3){1}}\put(45,45){\mbox{$q_f^+$}}
\put(-2,45){\mbox{\small time}}
\bezier{0}(20,21)(12,25)(20,29)\put(11,24){\mbox{$k_1^+$}}
\put(22,22){\mbox{$v$}}\put(43,38){\mbox{$w$}}
\end{picture}
\caption{$(2,\bar1,3,1)\times(e,\bar2,\bar3,f)$ \hspace{2cm} $(e,\bar2,\bar3,f)\times(2,\bar1,3,1)$\label{twoloop}}
\end{figure}
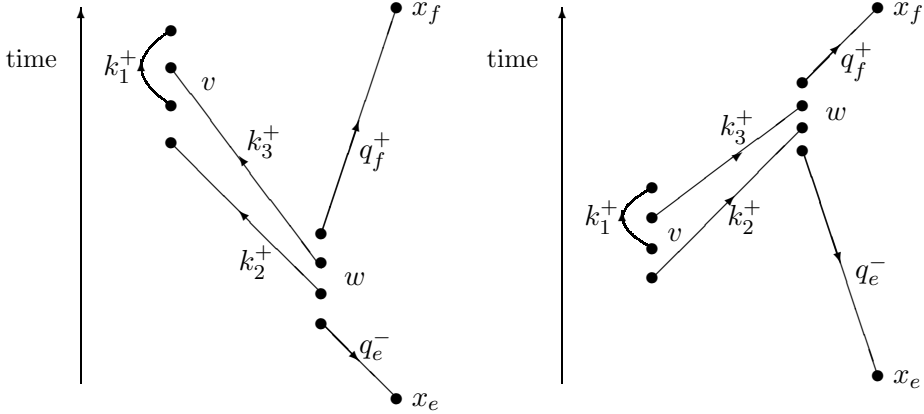
With $V=2$, $E=2$, $I=3$, eq.~(\ref{GammaE}) reads for the left graph, 
where the vertex $v$ is before the vertex $w$:
\begin{eqnarray}
&&\Gamma^{(2,2)} =\\&&
\quad\frac{g^2}{(4!)^2}\int \frac{d^3k_1d^3k_2d^3k_3}{(2\pi)^9
8\omega_1\omega_2\omega_3}\frac{i(2\pi)^3\delta^3(-\vec k_2
-\vec k_3 -\vec q_e-\vec q_f)}{-\omega_2-\omega_3+\omega_e-\omega_f+i
\varepsilon}\mathrm{exp}(i\theta^{\mu\nu}\phi_{\mu\nu}).\nonumber
\end{eqnarray}
We have $J_{v2}=J_{v3}=+1$, $J_{w2}=J_{w3}=-1$, $\sigma_e =+1$, $\sigma_f=
-1$. We obtain a non-trivial $I_{ij}$ term from the vertex $v$. For example,
the phase of the vertex $v$ in the left graph is
\begin{equation}
(2,\bar1,3,1)_v:  -k_1^+k_3^+  +\frac12k_2^+k_3^+,
\end{equation}
and is similar for the other 11 contributions.
For the vertex $w$ the $I_{ij}$ and $I_{ie}$ terms are non-zero. Again, we 
present only one contribution (note that $-q_e^- = +q_f^+$, owing to momentum 
conservation):
\begin{eqnarray}
(e,2,3,f)_w:&& \frac12 k_2^+(-q_e^-)+ \frac12 k_3^+(-q_e^-)+
\frac12 k_2^+q_f^+ + \frac12 k_3^+q_f^+ + \frac12 k_2^+k_3^+\nonumber\\
&&= \quad k_2^+q_f^+ + k_3^+q_f^+ + \frac12 k_2^+k_3^+.
\end{eqnarray}
Collecting the other 23 terms would be fairly edifying for a computer.
Summing up all contributions, using again $q_e^- = - q_f^+$,
integrating out $\vec k_3$ and setting 
$\varepsilon = 0$ yields
\begin{eqnarray}
\Gamma_{left}^{(2,2)} &=&
-\frac{ig^2}{(4!)^2}\int \frac{d^3k_2}{(2\pi)^3 8\omega_2^3}
\int\frac{d^3k_1}{(2\pi)^3 2\omega_1} 
2\cos \Big(\frac12k_2^+\tilde k_2^- \Big)
\nonumber \\
&&\quad\times \Big( 3 + e^{-i\theta^{\mu\nu}k_{1,\mu}^+ k_{2,\nu}^+}
+e^{+i\theta^{\mu\nu}k_{1,\mu}^+k_{2,\nu}^-}+
 e^{-i\theta^{\mu\nu}(k_{1,\mu}^+k_{2,\nu}^+
-k_{1,\mu}^+k_{2,\nu}^-)}\Big)\nonumber \\&&\quad
\times2\cos \Big(\frac12k_2^+\tilde k_2^- \Big)
\Big(6+2\cos(\theta^{\mu\nu}k_{2,\mu}^+q_{f,\nu}^+)
\nonumber\\&&\qquad+2\cos(\theta^{\mu\nu} k_{2,\mu}^-
q_{f,\nu}^+)+2\cos(\theta^{\mu\nu}
(k_{2,\mu}^+-k_{2,\mu}^-)q_{f,\nu}^+)\Big).
\end{eqnarray}
The first two lines of the integral kernel are exactly eq.~(\ref{tad}), 
with the obvious replacements (note the correct signs coming from 
the $\sigma$'s and $J$'s) $-q_e^- \rightarrow +k_2^+$ and 
$-q_f^+ \rightarrow +k_3^+\rightarrow - k_2^-$. 
For $\Gamma_{right}^{(2,2)}$ we find the same expression with $k_{2,3}^+
\rightarrow -k_{2,3}^+$, because of the reversed sign of $J_{ve}$, etc. 
This yields exactly the complex-conjugated expression, so
with the help of $4\cos^2(\frac x2) = 2+2\cos(x)$ we get 
\begin{eqnarray} \label{SnowmanTO2}
&&\Gamma_{tot}^{(2,2)} = -\frac{ig^2}{(4!)^2}\int \frac{d^3k_2}{(2\pi)^3 
8\omega_{2}^3}(2+2\cos(k_2^+\tilde k_2^-))\\&&\quad
\times\Big(6+2\cos(k_2^+\tilde q_f^+) 
+2\cos(k_2^-\tilde q_f^+) +2\cos((k_2^+-k_2^-)\tilde q_f^+)\Big)
\nonumber\\&&\times
\int\frac{d^3k_1}{(2\pi)^3 2\omega_{1}}\Big(6+2\cos(k_1^+\tilde k_2^+)+
2\cos(k_1^+\tilde k_2^-) +2\cos(k_1^+(\tilde k_2^+-\tilde k_2^-))\Big)
\nonumber.
\end{eqnarray}
Note the extra $i$ due to the slightly unusual definition of the 
S-matrix used in \cite{Noncom1}.

\subsection{Four-Point One-Loop Correction}      \label{Fishexactchapt}

\begin{figure}[h!] 
\begin{picture}(55,65)  
\put(19,34){\mbox{$\bullet$}}
\put(19,39){\mbox{$\bullet$}}
\put(19,44){\mbox{$\bullet$}}
\put(19,49){\mbox{$\bullet$}}
\put(39,10){\mbox{$\bullet$}}
\put(39,14){\mbox{$\bullet$}}
\put(39,18){\mbox{$\bullet$}}
\put(39,22){\mbox{$\bullet$}}
\put(40,11){\line(1,-1){10}}\put(40,11){\vector(1,-1){5}}
\put(40,23){\line(1,3){10}}\put(40,23){\vector(1,3){5}}
\put(49,0){\mbox{$\bullet$}}\put(52,0){\mbox{$x_e$}}
\put(49,52){\mbox{$\bullet$}}\put(52,52){\mbox{$x_f$}}
\put(20,35){\line(1,-1){20}}\put(29,18){\mbox{$k_2^+$}}
\put(20,45){\line(3,-4){20}}\put(30,34){\mbox{$k_3^+$}}
\put(5,0){\vector(0,1){60}}
\put(30,25){\vector(-1,1){1}}\put(45,7){\mbox{$q_e^-$}}
\put(30,32){\vector(-3,4){1}}\put(45,33){\mbox{$q_f^+$}}
\put(-5,55){\mbox{\small time}}
\put(20,40){\line(-1,-4){10}}\put(9,-1){\mbox{$\bullet$}}
\put(20,40){\vector(-1,-4){5}}\put(12,-1){\mbox{$x_g$}}
\put(20,50){\line(-1,1){10}}\put(9,59){\mbox{$\bullet$}}
\put(20,50){\vector(-1,1){5}}\put(12,59){\mbox{$x_h$}}
\put(10,20){\mbox{$q_g^-$}}
\put(11,52){\mbox{$q_h^+$}}
\put(24,42){\mbox{$v$}}\put(43,16){\mbox{$w$}}
\end{picture}
\qquad\qquad
\begin{picture}(55,65)  
\put(19,16){\mbox{$\bullet$}}
\put(19,20){\mbox{$\bullet$}}
\put(19,24){\mbox{$\bullet$}}
\put(19,28){\mbox{$\bullet$}}
\put(39,33){\mbox{$\bullet$}}
\put(39,36){\mbox{$\bullet$}}
\put(39,39){\mbox{$\bullet$}}
\put(39,42){\mbox{$\bullet$}}
\put(20,17){\line(1,1){20}}\put(40,43){\vector(1,1){5}}
\put(20,25){\line(4,3){20}}\put(40,34){\vector(1,-3){5}}
\put(49,3){\mbox{$\bullet$}}\put(52,3){\mbox{$x_e$}}
\put(49,52){\mbox{$\bullet$}}\put(52,52){\mbox{$x_f$}}
\put(40,43){\line(1,1){10}}\put(30,24){\mbox{$k_2^+$}}
\put(40,34){\line(1,-3){10}}\put(29,36){\mbox{$k_3^+$}}
\put(5,0){\vector(0,1){60}}
\put(30,27){\vector(1,1){1}}\put(47,17){\mbox{$q_e^-$}}
\put(31,33){\vector(4,3){1}}\put(45,45){\mbox{$q_f^+$}}
\put(-5,55){\mbox{\small time}}
\put(20,21){\line(-1,-1){10}}\put(9,10){\mbox{$\bullet$}}
\put(20,21){\vector(-1,-1){5}}\put(12,10){\mbox{$x_g$}}
\put(20,29){\line(-1,3){10}}\put(9,58){\mbox{$\bullet$}}
\put(20,29){\vector(-1,3){5}}\put(12,58){\mbox{$x_h$}}
\put(11,18){\mbox{$q_g^-$}}
\put(10,43){\mbox{$q_h^+$}}
\put(22,22){\mbox{$v$}}\put(43,38){\mbox{$w$}}
\end{picture}
\caption{Two contributions to $\Gamma^{(4,1)}$\label{fourpoint}}
\end{figure}
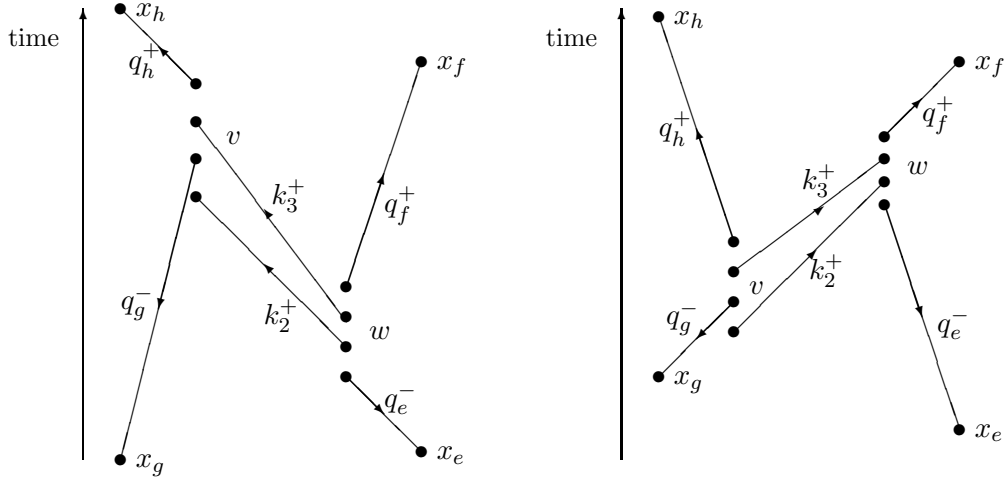

Finally, for the one-loop correction to the $t$-channel four-point function we have the 
contributions of figure \ref{fourpoint}.

Without going into detail with respect to the phase, we can prove the 
IR finiteness of the sum of these contributions:
\begin{eqnarray}
&&\Gamma^{(4,1)} = \frac{g^2}{(4!)^2}\int\frac{d^3k_2}{(2\pi)^32\omega_2}
\int\frac{d^3k_3}{(2\pi)^32\omega_3}\\&&\quad\times
\Big(\frac{i(2\pi)^3
\delta^3(-\vec k_2-\vec k_3-\vec q_e -\vec q_f)}{-\omega_2-\omega_3+\omega_e
-\omega_f+i\varepsilon}\nonumber\\&&\qquad\times
\Psi(-q_e^-,-q_f^+,-k_2^+,-k_3^+)
\Psi(-q_g^-,-q_h^+,k_2^+,k_3^+)
\nonumber\\&&+
\frac{i(2\pi)^3
\delta^3(-\vec k_2-\vec k_3-\vec q_g -\vec q_h)}{-\omega_2-\omega_3+\omega_g
-\omega_h+i\varepsilon}\nonumber\\&&\qquad\times
\Psi(-q_e^-,-q_f^+,+k_2^+,+k_3^+)
\Psi(-q_g^-,-q_h^+,-k_2^+,-k_3^+)
\Big).\nonumber
\end{eqnarray}
Here the phase $\Psi$ will be defined in section \ref{Vertex}.
With conservation of the global 4-momentum $\delta^4(q_e^-+q_f^++q_g^-+q_h^+)$,
we have $\omega_g-\omega_h = -(\omega_e-\omega_f)$ in the denominator
of the second term. Before integrating out $k_3$ we let $\vec k_2\rightarrow
-\vec k_2,\vec k_3\rightarrow-\vec k_3$ in the second term, so that 
$\vec k_3 = -\vec k_2-\vec q_e-\vec q_f$ in both terms. Thus we find
\begin{eqnarray}  \label{Fishexact}
&&\Gamma^{(4,1)} 
= -\frac{g^2}{(4!)^2}\int\frac{d^3k_2}{(2\pi)^32\omega_2}\frac{1}
{(2\pi)^32\omega_3}i(2\pi)^3\\&&\times
\Big(\frac{\Psi(-q_e^-,-q_f^+,-k_2^+,-k_3^+)
\Psi(-q_g^-,-q_h^+,k_2^+,k_3^+)}
{\omega_2+\omega_3-(\omega_e-\omega_f)-i\varepsilon}\nonumber\\&&+
\frac{\Psi(-q_e^-,-q_f^+,-k_2^-,-k_3^-)
\Psi(-q_g^-,-q_h^+,k_2^-,k_3^-)}
{\omega_2+\omega_3+(\omega_e-\omega_f)-i\varepsilon}\Big)
\bigg\arrowvert_{\vec k_3 = -(\vec k_2 + \vec q_e + \vec q_f)}.\nonumber
\end{eqnarray}
We find that the denominators are strictly positive,
\begin{eqnarray*}
&&|(\omega_2+\omega_3)|^2-|(\omega_e-\omega_f)|^2 =\\
&&\vec k_2^2+ m^2 + (\vec k_2 +\vec q_e +\vec q_f)^2 +m^2 +2\omega_2\omega_3
\\&&\hspace{2cm}-\vec q_e^2-m^2-\vec q_f^2 - m^2 +2\omega_e\omega_f =\\
&&2\Big(\vec k_2(\vec k_2 + \vec q_e + \vec q_f) + \vec q_e \vec q_f + 
\omega_2\omega_3 + \omega_e\omega_f\Big) \Big\arrowvert_{\vec k_3 = 
-(\vec k_2 + \vec q_e + \vec q_f)} >0 \\&&\hspace{2cm}
 (|\vec p\cdot\vec q| <\omega_p\omega_q,\ m> 0). 
\end{eqnarray*}
Thus, no new kinematic IR divergence occurs with respect to the commutative case, although 
the usual cancellations could not take place because of the different phases.
Hence we made sure that no novel problems arise from this quarter.

\section{The Feynman Rules for IPTOPT}  \label{Diagrammatics}

To obtain the set of diagrammatic rules for our model we have to answer three questions: 
\begin{enumerate}
\item What is the vertex?
\item What is the propagator?
\item How to construct graphs? 
\end{enumerate}
The first of these we postpone to section \ref{Vertex}, while the other two are tackled 
by retracing our steps to the explicit result for the tadpole obtained in \cite{Noncom1}.

\subsection{The Full Non-Commutative Propagator} \label{NcFullProp}

We start our search for the Feynman(-like) rules of non-commutative IPTOPT at the explicit expression
for the two-point one-loop tadpole $G^{(2,1)}$, eq.~(24) of \cite{Noncom1}.

Repeating the notation from \cite{Noncom1} (recall that $p^\pm:= (\pm \omega_p,
\vec p),\ \omega_p:= \sqrt{\vec p^2+m^2}, \ \tilde p^\nu:= p_\mu\theta^{\mu\nu})$:
\begin{eqnarray}
\mathcal{I}^{\pm\pm}((\pm p)^+,(\pm q)^+) &=&   
\int \frac{d^3 k}{(2\pi)^3 2 \omega_k}\,
\big( 3 
+ \mathrm{e}^{i p^\pm \tilde{k}^+ 
+ i q^\pm \tilde{k}^+ }
+ \mathrm{e}^{i p^\pm \tilde{k}^+ }
+ \mathrm{e}^{i q^\pm \tilde{k}^+ }\big) \nonumber\\ &\equiv&
\mathcal{I}(p^\pm,q^\pm) \:,
\label{Ipq}
\end{eqnarray}
we retrace one step and give the unamputated FT Green function
\begin{align} \label{G21full}
G^{(2,1)}(p,q) & =
 - \lim_{\delta_1,\delta_2 \to 0} \frac{g}{12} (2\pi)^4 \delta(p+q)\,
\nonumber \\
& \times \Big( \frac{1}{p_0{-}\omega_p{+}i\delta_1} \,\frac{1}{\omega_p{+}\omega_q{-}i\delta_2} 
\frac{\cos(\tfrac{1}{2} p^+ \tilde{q}^+)}{4 \omega_p \omega_q}\mathcal{I}(p^+,q^+)\nonumber \\
& \quad + \frac{1}{q_0{-}\omega_q{+}i\delta_1} \,\frac{1}{\omega_p{+}\omega_q{-}i\delta_2} 
\frac{\cos(\tfrac{1}{2} p^+ \tilde{q}^+) }{4 \omega_p \omega_q} \mathcal{I}(p^+,q^+) \nonumber \\
& \quad + \frac{1}{p_0{-}\omega_p{+}i\delta_1} \, \frac{1}{q_0{+}\omega_q{-}i\delta_2} 
\frac{\cos(\tfrac{1}{2} p^+ \tilde{q}^-) }{4 \omega_p \omega_q} \mathcal{I}(p^+,q^-) \nonumber \\
& \quad + \frac{1}{q_0{-}\omega_q{+}i\delta_1} \, \frac{1}{p_0{+}\omega_p{-}i\delta_2} 
\frac{\cos(\tfrac{1}{2} p^- \tilde{q}^+) }{4 \omega_p \omega_q} \mathcal{I}(p^-,q^+) \nonumber \\
& \quad + \frac{1}{\omega_p {+}\omega_q{-}i\delta_1} \, \frac{1}{{-}q_0{-}\omega_q{+}i\delta_2} 
\frac{\cos(\tfrac{1}{2} p^- \tilde{q}^-) }{4 \omega_p \omega_q} \mathcal{I}(p^-,q^-) \nonumber \\
& \quad + \frac{1}{\omega_p {+}\omega_q{-}i\delta_1} \, \frac{1}{{-}p_0{-}\omega_p{+}i\delta_2} 
\frac{\cos(\tfrac{1}{2} p^- \tilde{q}^-) }{4 \omega_p \omega_q} \mathcal{I}(p^-,q^-) \Big)\;.
\end{align}

Making use of local energy--momentum conservation 
$p^0=-q^0$ and $\omega_p=+\omega_q$, 
and of the relation $q^\pm=-p^\mp$, we eliminate $q$ and contract eq. (\ref{G21full}) to
\begin{eqnarray}
& & G^{(2,1)}(p) = -\lim_{\varepsilon \rightarrow 0}
\frac{g}{12(2\omega_p)^2} (2\pi)^4 \delta(p+q) \nonumber \\
& & \times \Big( \frac{1}{p_0-\omega_p+i \varepsilon}\  \frac{1}{p_0+\omega_p-i \varepsilon} \  
          \cos \Big( \frac{1}{2}p^+\tilde{p}^- \Big) \big( \mathcal{I}(p^+,-p^-)+ \mathcal{I}(p^-,-p^+) \big) \nonumber \\
& & \quad -\frac{1}{p_0-\omega_p+i \varepsilon}\   \frac{1}{p_0-\omega_p+i \varepsilon}\ 
          \cos \Big( \frac{1}{2}p^+\tilde{p}^+ \Big)\ \mathcal{I}(p^+,-p^+)  \nonumber \\
& & \quad - \frac{1}{p_0+\omega_p-i \varepsilon} \  \frac{1}{p_0+\omega_p-i \varepsilon} \ 
          \cos \Big(\frac{1}{2}p^-\tilde{p}^- \Big)\ \mathcal{I}(p^-,-p^-) \Big).
\end{eqnarray}
This can easily be written as the sum over two signs:
 \begin{eqnarray} \label{G21}
G^{(2,1)}(p) & = & \frac{g}{12} (2\pi)^4 \delta(p+q) \sum_{\sigma}^{+1,-1} \sum _{\sigma'}^{+1,-1} 
         \cos(p^{\sigma} \tilde{p}^{\sigma'}) \mathcal{I}(p^{\sigma},-p^{\sigma'})\nonumber \\
      & &     \frac{1}{2 \omega_p} \frac{1}{\sigma p_0 - \omega_p + i \varepsilon} \quad
            \frac{1}{2 \omega_p} \frac{1}{\sigma' p_0 - \omega_p + i \varepsilon}.
\end{eqnarray}

\subsection{The TO Propagator}

Equation~(\ref{G21}) lets us read off the answers to both our questions. Since we have not performed any amputation yet, 
two propagators must be included in the above expression. We easily identify the TO propagator as 
\begin{equation} \label{Propagator} 
  i \Delta^{TO} := \frac{\delta_{\sigma, -\sigma'}}{2 \omega_p} 
            \frac{i}{\sigma p^0 - \omega_p + i \varepsilon}.
\end{equation}
The $\delta_{\sigma, -\sigma'}$ was included to guarantee TO-diagrammatic consistency: every directed TO line that 
leaves one vertex ($\sigma$) has to arrive at another one ($\sigma'$). (The correctness of this addition will
become evident in the following examples.)

Note that the same result is independently obtained in \cite{Denk}, 
where the TO propagator is called ``contractor''.

The global TO of the vertices is another necessary issue to be encoded in $\Delta^{TO}$: every line
has to leave its earlier vertex and arrive at its later vertex, and this must be consistently so 
for all lines of the diagram. 
This property is taken care of by the sign of the pole prescription.
As illustrated in the amplitudes (re)calculated in section \ref{IPTOPTFR}, only products of TO propagators 
in TO consistent graphs (if $A < B$ and $C < A$ then $C < B$) will contribute. 
All others (e.g. $A < B$ and $C < A$ but $B < C$) will have their poles 
bundled in the same complex half-plane and hence vanish upon 
integrating over $p^0$.

\subsection{Building Graphs}

In addition to providing us with a propagator, eq.~(\ref{G21}) also tells us how to construct graphs:
multiply together all the building blocks for a graph of given topology --- lines, vertices, subgraphs --- 
which all depend on the entering or leaving ($\sigma_i = \pm 1$) of the lines running into them.
Then sum over all signs.
The propagators take care of the correct connection of all parts of the diagram, especially causal consistency:
if vertex A is later than vertex B and B is later than C, than A is also later than C.

Even at this point we may already calculate the two-point zero-loop function, the usual covariant propagator,
\begin{eqnarray}  
i \Delta_F & = &  \sum_{\sigma}^{+1,-1} i \Delta^{TO}(\sigma)   
   = \sum_{\sigma}^{+1,-1}  \frac{1}{2 \omega} \frac{i}{\sigma p_0 - \omega +i \varepsilon} \nonumber \\
   & = & \frac{i}{2 \omega} \Big( \frac{1}{+p_0 - \omega + i \varepsilon } + \frac{1}{-p_0 - \omega + i \varepsilon} \Big) 
    = \frac{i}{ p_0^2 - \omega^2 + i \varepsilon }. 
\end{eqnarray}

\subsection{Complete One-Loop Integrals}

To complete our discussion of $G^{(2,1)}$, and for further use in section \ref{MouseFR}, we evaluate
the  $\mathcal{I}$'s occurring in eq.~(\ref{G21}).

Abbreviating the (cut-off-regularized) divergent part of the planar term
by $\mathcal{Q} = \Lambda^2+\frac{m^2}{2}\ln(\frac{m^2}{\Lambda^2})$, we give 
$\mathcal{I}(p^+,-p^+)$, which was already calculated in \cite{Noncom1}, eq.~(31):
\begin{equation}   \label{I++} 
\mathcal{I}(p^+,-p^+) = \frac{2}{(2\pi)^2} 
\Big( \mathcal{Q} - \sqrt{-\frac{m^2}{\tilde{p}^2_+}} 
K_1\Big(\sqrt{-m^2 \tilde{p}^2_+}\Big) \Big).
\end{equation}
Analogously we find
\begin{equation} \label{I-+}
\mathcal{I}(p^-,-p^-) = \frac{2}{(2\pi)^2}
\Big( \mathcal{Q} -\sqrt{-\frac{m^2}{\tilde{p}^2_-}}K_1\Big(\sqrt{-m^2\tilde{p}^2_-}\Big)\Big).
\end{equation}
Calculating the sum of the remaining integrals still has to be done.
Adding the integrands gives 
\begin{eqnarray} 
&&\mathcal{I}(p^+,-p^-) + \mathcal{I}(p^-,-p^+) =   \nonumber\\&&
=\int \frac{d^3 k}{(2\pi)^3 2 \omega_k}\,
\big( 6+
\mathrm{e}^{- ik^+\tilde{p}^+ +ik^+\tilde{p}^- }
+ \mathrm{e}^{- ik^+ \tilde{p}^+  }
+ \mathrm{e}^{+ ik^+\tilde{p}^-  }\nonumber\\&&\qquad\qquad\qquad
\qquad
+ \mathrm{e}^{+ ik^+\tilde{p}^+ - ik^+ \tilde{p}^- }
+ \mathrm{e}^{+ ik^+\tilde{p}^+  }
+ \mathrm{e}^{- ik^+\tilde{p}^-  }
\big) 
\\&&
= \int \frac{d^3 k}{(2\pi)^3 2 \omega_k}\, 2 \big(3+\cos(k^+\tilde{p}^+) 
+\cos(k^+ \tilde{p}^-)
 +\cos( k^+ (\tilde{p}^+ - \tilde{p}^-)) \big).\nonumber
\end{eqnarray}
The first and second cosine terms are just the ones yielding the 
non-planar parts of eqs.~(\ref{I++}) and (\ref{I-+}). 
The third one has to be dealt with explicitly. With 
$(\tilde{p}^+ - \tilde{p}^-)_\mu = 2 \Theta_{0\mu} \omega$ and $\theta_{00}
=0$ we can choose a coordinate system with the $z$-axis parallel to the 
3-vector $\theta_{0i}$. Thus 
integrating out the angles yields  
\begin{equation}
\frac{2}{(2\pi)^2 |\Theta_{0i}| \omega} \int_0^\infty dk
\frac{|\vec k|}{\omega_k} \sin(2 |\vec k||\Theta_{0i}| \omega).
\end{equation}
This we evaluate as
\begin{equation} \label{I++ + I--}
= \frac{m}{(2\pi)^2 |\Theta_{0i}|\omega}K_1\big(2m|\Theta_{0i}|\omega\big).
\end{equation}
Hence we have 
\begin{eqnarray}
& & \mathcal{I}(p^+,-p^-) + \mathcal{I}(p^-,-p^+) = \frac{2}{(2\pi)^2} 
    \Big( \frac32 \mathcal{Q} -  2 \frac{m}{|\tilde{p}^+|} K_1(m|\tilde{p}^+|) \nonumber \\ 
 && - 2  \frac{m}{|\tilde{p}^-|} K_1(m|\tilde{p}^-|)  
      + \frac{m}{ |\Theta_{0i}| \omega} K_1\big( 2 m|\Theta_{0i}|\omega \big)\Big).
\end{eqnarray}

For further use (eq.~(\ref{Mouse1})), we finally present another result.
Iff $\mathcal{I}(p^+\!,\!-p^-)$ occurs under an integral over $d^3p$ together with functions $f(\vec{p})$ 
invariant under $\vec{p} \to -\vec{p}$  we have:
\begin{align}  \label{Ineu} 
&\int d^3p f(\vec{p}) \mathcal{I}(p^+,-p^-) = \int d^3p f(\vec{p})\frac{1}{(2 \pi)^2}\nonumber\\&\quad\times 
    \Big( \frac32 \mathcal{Q} - \frac{2 m}{|\tilde{p}^+|} K_1(m  |\tilde{p}^+|)
                        - \frac{2 m}{|\tilde{p}^-|} K_1(m  |\tilde{p}^-|)
                        + \frac{m}{\omega_p |\Theta_{0i}|} K_1(2 m \omega_p |\Theta_{0i}|)  \Big).
\end{align}
$\mathcal{I}(p^-,-p^+)$ yields an identical result under the same assumption.

\subsection{The Vertex} \label{Vertex}

To answer our first question we straightforwardly peruse eq.~(\ref{GammaE}) for no internal lines and 
four external ones with general causalities ($\sigma$'s). 
Summing over all possible inner (nano-) TO of the vertex, we proceed as in section 
\ref{Examples} and find ($\tilde p^\nu := p_\mu\theta^{\mu\nu}$)
\begin{align} \label{VertexFR}
&\Gamma^{(4,0)}(p^{\sigma_1}_1 ,p^{\sigma_2}_2 ,p^{\sigma_3}_3 ,p^{\sigma_4}_4  ) :=
   \frac{g}{4!} \Psi(-p^{\sigma_1}_1 ,-p^{\sigma_2}_2 ,-p^{\sigma_3}_3 ,-p^{\sigma_4}_4 ) = \nonumber \\&
  = \frac{g}{3} \Big( \cos \Big(\frac12p^{\sigma_1}_{1}\tilde p^{\sigma_2}_{2}\Big)
\cos \Big(\frac12 p^{\sigma_3}_{3}\tilde p^{\sigma_4}_{4} \Big)
\cos \Big(\frac12 (p^{\sigma_1}_{1}+p^{\sigma_2}_{2})(\tilde p^{\sigma_3}_{3}
+\tilde p^{\sigma_4}_{4})\Big)
\nonumber\\
&+ (2) \leftrightarrow (3) + (2) \leftrightarrow (4) \Big).
\end{align}
Note that here all the momenta are defined outgoing of the vertex. With the 
symmetry of the cosine we explicitly check the invariance of (\ref{VertexFR}) 
with respect to any permutation of the momenta.

Unfortunately, the tadpole has to be treated separately. From 
eq.~(\ref{GammaE})
it follows that the tadpole line has to be oriented forward in time. 
Thus only $\frac{24!}{2}$ nano-configurations at the vertex contribute.
We find for the phase factor of a 1-loop tadpole (defining $p_2^{\sigma_2},\ p_3^{\sigma_3}$ outgoing, loop momentum $p_1^+$)
\begin{eqnarray} \label{TadpolVertexFR}
&&\frac g{4!}\exp\Big(i\theta_{\mu\nu}\sum_{a,b=1}^3\tau^v_{ab}p_a^{\sigma_a}
p_b^{\sigma_b}\Big)=:\frac g{4!}
\Phi(p_1^+;-p_{2}^{\sigma_2},-p_{3}^{\sigma_3})\nonumber\\&&
= \frac g{12}\Big(3+e^{i p^+_{1}\tilde p_{2}^{\sigma_2}}+
e^{i p^+_{1}\tilde p_{3}^{\sigma_3}}+
e^{i p^+_{1}(\tilde p_{2}^{\sigma_2}+\tilde p_{3}^{\sigma_3})}
\Big)\cos \Big(\frac12 p^{\sigma_2}_{2}\tilde
p_{3}^{\sigma_3}\Big).
\end{eqnarray}

\subsection{Summary of Diagrammatics}

To calculate a Fourier-transformed, amputated amplitude, use the following rules: 
\begin{enumerate}
   \item An amputated external line carries the momentum $q_e^{\sigma_e}$; 
$\sigma_e=+1$ if the line is directed into the future, 
$\sigma_e=-1$ if it runs into the past:
\begin{equation}
q_e^{\sigma_e} = (\sigma_e \sqrt{\vec{q}^2 + m^2}, \vec{q})^T.
\end{equation}
 \item For a general, non-tadpolic vertex write a factor
\begin{eqnarray}
& &\frac g{4!}\Psi(-p_1^{\sigma_1},-p_2^{\sigma_2},-p_3^{\sigma_3},-p_4^{\sigma_4})= \nonumber\\ 
& =& \frac{g}{3} \Big( \cos\Big(\frac12p^{\sigma_1}_{1}\tilde p^{\sigma_2}_{2}\Big)
\cos\Big(\frac12 p^{\sigma_3}_{3}\tilde p^{\sigma_4}_{4}\Big)
\cos\Big(\frac12 (p^{\sigma_1}_{1}+p^{\sigma_2}_{2})(\tilde p^{\sigma_3}_{3}
+\tilde p^{\sigma_4}_{4})\Big) \nonumber \\
&+& (2) \leftrightarrow (3) + (2) \leftrightarrow (4) \Big),
\end{eqnarray}
where all momenta are oriented outwards from the vertex.
   \item For a tadpolic vertex (with loop momentum $p_1^+$),
write a factor
\begin{eqnarray}
& & \frac g{4!}\Phi(p_1^+;-p_2^{\sigma_2},-p_3^{\sigma_3})= \nonumber \\ 
&=& \frac g{12}\Big(3+e^{i p^+_{1}\tilde p_{2}^{\sigma_2}}+
e^{i p^+_{1}\tilde p_{3}^{\sigma_3}}+ e^{i p^+_{1}(\tilde p_{2}^{\sigma_2}+\tilde p_{3}^{\sigma_3})}
\Big) \cos\Big(\frac12 p^{\sigma_2}_{2}\tilde p_{3}^{\sigma_3}\Big),
\end{eqnarray}
where $p_2, p_3$ are oriented outwards from the vertex.
    \item For an inner line, write the propagator 
\begin{equation}
i \Delta^{TO} = \frac{i}{2 \omega} \frac{\delta_{\sigma, -\sigma'}}{\sigma p^0 - \omega_p + i \varepsilon}.
\end{equation}
    \item Sum over all $\sigma$'s of the internal lines in order to include all 
possible
contributions with respect to the time ordering of the inner vertices.   
    \item Integrate over all loop momenta (including tadpole momenta).
\end{enumerate}
Remember that 4-momentum conservation is valid at all vertices and along all lines.

\section{Examples for the Application of the Feynman Rules for NC-IPTOPT}   \label{IPTOPTFR}

In order to both illustrate the applicability and demonstrate the validity 
of the new-found FR (and since a motivation was given for them, rather than a derivation), 
we employ them in the recalculation of the diagrams of section \ref{Examples}.

In addition we will finally be able to calculate the ``mouse''-diagram $\Gamma^{(2,3)}$, which was one of the main 
motivations for the development of this diagrammatics.

\subsection{The Diagrammatic Tadpole}     \label{TadpoleFR}


Once again we turn toward the tadpole, obtained by explicitly commuting out the free-field operators
in \cite{Noncom1} and by use of the IPTOPT formula eq.~(39), \emph{ibidem}.

Simplifying eq.~(\ref{TadpolVertexFR}) by using 4-momentum conservation $p_2^{\mu}=: q^{\mu} = -p_3^{\mu}$, 
setting the external momenta on-shell $\sigma_2=+1,\sigma_3=-1$, and defining $p_1^+ =: k^+,\ 
p_\mu\Theta^{\mu \nu}=: \tilde{p}$ we 
find for the vertex factor
\begin{equation} \label{TadVert2}
\frac{g}{4!} \Phi(k^+;-q^+,q^+) = \frac{g}{6} \big(2+\cos(k^+\tilde{q}^+)\big).
\end{equation}
Note that the $\sigma$ of the looped line does not occur. 
Multiplying with the propagator eq.~(\ref{Propagator}), summing over $\sigma, \sigma'$ and integrating over 
phase space then yields the FT NC tadpole amplitude, which is well known by now: 
\begin{eqnarray} \label{Tadpole}
 \Gamma^{(2,1)}&=&\int \frac{d^4k}{(2\pi)^4}\frac{g}{6} (2+\cos(k^+\tilde{q}^+)) \sum_{\sigma, \sigma'}^{\pm 1}
        \frac{\delta_{\sigma, -\sigma'}}{2 \omega} \frac{i}{\sigma k^0 - \omega_k + i \varepsilon} \nonumber \\
     &=&  \int \frac{d^4k}{(2\pi)^4}\frac{g}{6} (2+\cos(k^+\tilde{q}^+))\frac{i}{2 \omega_k} 
          \Big( \frac{1}{k^0-\omega_k+i\varepsilon} - \frac{1}{k^0+\omega_k-i\varepsilon} \Big) \nonumber \\
     &=&  \frac{g}{6} \int \frac{d^3k}{(2\pi)^3}\frac{1}{2 \omega_k} (2+\cos(k^+\tilde{q}^+)).
\end{eqnarray}
The actual $k^0$-integration can be performed directly for both terms separately, heeding 
non-vanishing semicircles at infinity.
Alternatively they can be brought over a common denominator, resulting 
in the usual Feynman propagator.

\subsection{The Diagrammatic Snowman} \label{SnowmanFR}


To further strengthen our confidence in $\Delta^{TO}$ and the vertices of eqs.~(\ref{VertexFR}) and (\ref{TadpolVertexFR}),
we demonstrate how to utilize them to evaluate the snowman of section \ref{Snowmanexact}.



To obtain the amputated, FT snowman amplitude, we multiply the terms for the two
vertices with each other and with one $\Delta^{TO}$ for the head-loop and two for the body-loop. 
Using 4-momentum conservation $k_3^{\mu}=-k_2^{\mu}, k_3^\pm=-k_2^\mp$, summing over 
$\sigma_1^v,\sigma_1^w,\sigma_2^v, \sigma_2^w,\sigma_3^v,\sigma_3^w = \pm1$ 
and integrating over the two loop-momenta $k_1^{\mu}, k_2^{\mu}$, we find

\begin{eqnarray}
&&\Gamma^{(2,2)} =\frac{g}{4!}\int \frac{d^4k_1}{(2 \pi)^4}\nonumber\\&&
 \quad\sum_{\sigma_2^v,\sigma_2^w,\sigma_3^v,\sigma_3^w}^{+,-}
                   \overbrace{\frac{g}{4!} \int \frac{d^4k_2}{(2 \pi)^4}\sum_{\sigma_1, \sigma'_1} \frac{i}{2\omega_1}
                   \frac{\delta_{\sigma_1,-\sigma'_1}}{\sigma_1 k_1^0-\omega_1+i \varepsilon} 
                   \Phi^v(k_1^+;-k_2^{\sigma_2^v}, +k_2^{-\sigma_3^v})}^{\Gamma^{(2,1)}(k_2; \sigma_2^v, \sigma_3^v)} \nonumber \\   
 & &\hspace{2cm}\times
 \Psi^w(-q_e^-,-q_f^+, +k_2^{-\sigma_2^w}, -k_2^{+\sigma_3^w}) 
          \nonumber\\&&\hspace{2cm}\times  
\frac{i}{2\omega_2} \frac{i}{2\omega_2}        
            \quad  \frac{\delta_{\sigma_2^v,-\sigma_2^w}}{\sigma_2^v k_2^0-\omega_2 + i \varepsilon}  
            \quad   \frac{\delta_{\sigma_3^v,-\sigma_3^w}}{-\sigma_3^v k_2^0-\omega_2 + i \varepsilon} \nonumber \\
    && = \frac{i g^2}{(4!)^2}\int \frac{d^4k_1}{(2 \pi)^4}\frac{d^4k_2}{(2 \omega_2)^2(2 \pi)^4}\nonumber\\&&\qquad
                   \sum_{\sigma_2^v,\sigma_3^v}^{+,-} 
                   \Phi^v(k_1^+;-k_2^{\sigma_2^v}, +k_3^{-\sigma_3^v})\Psi^w(-q_e^-,-q_f^+, +k_2^{\sigma_2^v}, -k_2^{-\sigma_3^v})  
\nonumber \\
          & &\quad\times\ \frac{1}{(k_1^0)^2-\omega_1^2+i \varepsilon}  
            \   \frac{1}{\sigma_2^v k_2^0-\omega_2 + i \varepsilon}  
            \   \frac{1}{\sigma_3^v k_2^0+\omega_2 - i \varepsilon} \nonumber \\
   && = \frac{g^2}{(4!)^2}\int \frac{d^3k_1}{2\omega_1(2 \pi)^3}\frac{d^3k_2}{(2 \omega_2)^2(2 \pi)^3} \frac{dk^0_2}{2\pi}
                 \nonumber \\
       && \quad\Big( \Phi^v(k_1^+;-k_2^+, +k_2^-) \Psi^w(-q_e^-,-q_f^+, +k_2^+, -k_2^-) 
        \nonumber\\&&\hspace{1.7cm}\qquad\times\quad \frac{1}{+k_2^0-\omega_2 + i \varepsilon} 
                 \   \frac{1}{+k_2^0+\omega_2 - i \varepsilon} \nonumber \\
       &&\quad +   \Phi^v(k_1^+;-k_2^+,+k_2^+)\Psi^w(-q_e^-,-q_f^+, +k_2^+, -k_2^+)
        \nonumber\\&&\hspace{1.7cm}\qquad\times\quad\frac{1}{+k_2^0-\omega_2 + i \varepsilon} 
                 \   \frac{1}{-k_2^0+\omega_2 - i \varepsilon} \nonumber \\
       && \quad+   \Phi^v(k_1^+;-k_2^-,+k_2^-)\Psi^w(-q_e^-,-q_f^+, +k_2^-, - k_2^-)
        \nonumber\\&&\hspace{1.7cm}\qquad\times\quad\frac{1}{-k_2^0-\omega_2 + i \varepsilon} 
                 \   \frac{1}{+k_2^0+\omega_2 - i \varepsilon} \nonumber \\
       &&\quad +   \Phi^v(k_1^+;-k_2^-,+k_2^+)\Psi^w(-q_e^-,-q_f^+, +k_2^-, -k_2^+) 
        \nonumber\\&&\hspace{1.7cm}\qquad\times\quad\frac{1}{-k_2^0-\omega_2 + i \varepsilon} 
                 \   \frac{1}{-k_2^0+\omega_2 - i \varepsilon} \Big).
\end{eqnarray}
In the last step we integrated over $k_1^0$ as in section \ref{TadpoleFR} and expanded the sums over $\sigma_2^v, \sigma_3^a$. 

Performing the $k^0_2$ integration reveals how $\Delta^{TO}$ selects the correct $\sigma$-signs: 
the poles in the second and the third term are double poles, both lying on top of each other in the same complex half-plane.
Hence we may close the contour in the other half without enclosing any residuum, yielding a vanishing integral
(mark that the auxiliary semicircle is harmless, contrary to the tadpole case).

In the first and the fourth term the poles lie in opposite halves and yield upon integration $2\pi i / (2 \omega_2)$.
Hence we find
\begin{eqnarray} \label{Snowman2FR}
 \Gamma^{(2,2)} &=& -\frac{i g^2}{(4!)^2}\int \frac{d^3k_1}{2\omega_1(2 \pi)^3}\frac{d^3k_2}{(2 \omega_2)^3(2 \pi)^3}
        \\ & & \Big( \Phi(k_1^+;-k_2^+, +k_2^-) \Psi(-q_e^-,-q_f^+, +k_2^+, -k_2^-) + \nonumber\\&&\quad\Phi(k_1^+;-k_2^-, +k_2^+)
 \Psi(-q_e^-,-q_f^+, +k_2^-, -k_2^+) \Big).\nonumber
\end{eqnarray}
Evaluation of the phases $\Phi$ and $\Psi$, using momentum conservation $q_e^-= -q_f^+$ 
and doing some trivial but tedious trigonometry, yields
\begin{eqnarray*}
&&\Phi(k_1^+;-k_2^+, +k_2^-) =\\&&\qquad 2 \cos(\frac12 k_2^+\tilde  k_2^-) 
\big( 3 + e^{-i k_1^+ \tilde{k}_2^+} +  
             e^{+i k_1^+ \tilde{k}_2^-}  +  e^{-ik_1^+ (\tilde{k}_2^+ - \tilde{k}_2^- )}   \big)  \\
&&\Phi(k_1^+;-k_2^-, +k_2^+) =\\&&\qquad 2 \cos(\frac12 k_2^+\tilde  k_2^-) 
\big( 3 + e^{+i k_1^+ \tilde{k}_2^+} +  
             e^{-i k_1^+\tilde{k}_2^-}  +  e^{+i k_1^+(\tilde{k}_2^+ - \tilde{k}_2^-)}  \big)  \\
&&\Psi(-q_e^-,-q_f^+, +k_2^+, -k_2^-) = 4 \cos(\frac12 k_2^+\tilde  k_2^-) \big( 3 + \cos(k_2^+\tilde  q_f^+) +  \nonumber \\
      & &\hspace{4cm}\qquad \cos(k_2^-\tilde  q_f^+) + \cos(( k_2^+ - k_2^-)\tilde  q_f^+ ) \big) \\
&&\Psi(-q_e^-,-q_f^+, +k_2^-, -k_2^+) =   \Psi(-q_e^-,-q_f^+, -k_2^+, +k_2^-).     
\end{eqnarray*}
Inserting this into eq.~(\ref{Snowman2FR}) we find exactly the same result as eq.~(\ref{SnowmanTO2}), as obtained by the TO procedure 
in section \ref{Snowmanexact}.

\subsection{The Diagrammatic Fish}

To demonstrate that our diagrammatic rules also work in a non-tadpolic context, we recalculate 
the $t$-channel four-point one-loop fish graph evaluated in section \ref{Fishexactchapt}. As above we restrict ourselves to the 
$t$-channel.

Using the same notation as in fig.~\ref{fourpoint}, we fix the external on-shell momenta as 
above: $\sigma_e= -1,\sigma_f=+1 ,\sigma_g=-1 ,\sigma_h=+1$.
4-momentum conservation yields
\begin{equation}  \label{momentumconservationfish}
\vec{k}_3 = -\vec{k}_2-\vec{q}_e-\vec{q}_f, \qquad 
k_3^0 = -k_2^0 +\omega_e -\omega_f \quad. 
\end{equation}
$\Gamma^{(4,1)}$ is then given as
\begin{eqnarray}   \label{FishFR1}
\Gamma^{(4,1)} &=& \frac{g^2}{(4!)^2} \int \frac{d^4k_2}{(2\pi)^4} \sum^{+1,-1}_{\sigma_2^w,\sigma_2^v,\sigma_3^w, \sigma_3^v}
              \frac{i}{2\omega_2}\frac{i}{2\omega_3} \nonumber \\ & &
              \Psi^w(-q_e^-,-q_f^+, -k_2^{\sigma_2^w}, -k_3^{\sigma_3^w}) 
              \Psi^v(-q_g^-,-q_h^+, +k_2^{-\sigma_2^v}, +k_3^{-\sigma_3^v})\nonumber \\ & &
              \frac{\delta_{\sigma_2^w, -\sigma_2^v}}{\sigma_2^w k_2^0-\omega_2+i\varepsilon}
              \quad \frac{\delta_{\sigma_3^w, -\sigma_3^v}}{-\sigma_3^w (k_2^0+\omega_e -\omega_f)-\omega_3+i\varepsilon} \nonumber \\
     &=&      \frac{g^2}{(4!)^2} \int \frac{d^4k_2}{(2\pi)^4} \sum^{+1,-1}_{\sigma_2^w,\sigma_3^w}
              \frac{1}{ 2\omega_2 2\omega_3} \nonumber \\ & &
              \Psi^w(-q_e^-,-q_f^+, -k_2^{\sigma_2^w}, -k_3^{\sigma_3^w}) 
              \Psi^v(-q_g^-,-q_h^+, +k_2^{+\sigma_2^w}, +k_3^{+\sigma_3^w})\nonumber \\ & &
              \frac{1}{\sigma_2^w k_2^0-\omega_2+i\varepsilon} \quad 
              \frac{1}{\sigma_3^w k_2^0+\sigma_3^w \omega_e -\sigma_3^w \omega_f+\omega_3-i\varepsilon} \nonumber \\
 &=&      \frac{g^2}{(4!)^2} \int \frac{d^4k_2}{(2\pi)^4} \frac{1}{2\omega_2} \frac{1}{2\omega_3}\nonumber \\ & &
  \Big( \frac{\Psi^w(-q_e^-,-q_f^+, -k_2^+, -k_3^+)}{k_2^0-\omega_2+i\varepsilon} 
        \frac{\Psi^v(-q_g^-,-q_h^+, +k_2^+, +k_3^+)}{k_2^0+\omega_e-\omega_f+\omega_3-i\varepsilon}
                          \nonumber \\ 
&+& \frac{\Psi^w(-q_e^-,-q_e^+, -k_2^+, -k_3^-)}{+k_2^0-\omega_2+i\varepsilon} 
        \frac{\Psi^v(-q_g^-,-q_h^+, +k_2^+, +k_3^-)}{-k_2^0-\omega_e+\omega_f+\omega_3-i\varepsilon}
                          \nonumber \\  
&+& \frac{\Psi^w(-q_e^-,-q_f^+, -k_2^-, -k_3^+)}{-k_2^0-\omega_2+i\varepsilon} 
        \frac{\Psi^v(-q_g^-,-q_h^+, +k_2^-, +k_3^+)}{+k_2^0+\omega_e-\omega_f+\omega_3-i\varepsilon}
                          \nonumber \\ 
&+& \frac{\Psi^w(-q_e^-,-q_f^+, -k_2^-, -k_3^-)}{-k_2^0-\omega_2+i\varepsilon} 
        \frac{\Psi^v(-q_g^-,-q_h^+,+k_2^-, +k_3^-)}{-k_2^0-\omega_e+\omega_f+\omega_3-i\varepsilon}
 \Big).
\end{eqnarray}
Inspecting the complex $k^0_2$ plane of the four terms we see that the poles of the second and the third term 
lie on the same half-plane and hence yield vanishing integrals. We thus find (for shortness we retain $k_3, \omega_3$, 
but of course eqs.~(\ref{momentumconservationfish}) still apply)
\begin{eqnarray}   \label{FishFR2}
\Gamma^{(4,1)} &=& \frac{-i g^2}{(4!)^2} \int \frac{d^3k_2}{(2\pi)^3} \frac{1}{2\omega_2} \frac{1}{2\omega_3} \\
   & & \Big(  \frac{\Psi^w(-q_e^-,-q_f^+, -k_2^+, -k_3^+) 
                    \Psi^v(-q_g^-,-q_h^+, +k_2^+, +k_3^+)}{\omega_2+\omega_3+\omega_e-\omega_f-i\varepsilon}
                          \nonumber \\ 
  && + \frac{\Psi^w(-q_e^-,-q_f^+, -k_2^-, -k_3^-)
             \Psi^v(-q_g^-,-q_h^+,+k_2^-, +k_3^-)}{\omega_2+\omega_3-(\omega_e-\omega_f)-i\varepsilon}
 \Big) \nonumber,
\end{eqnarray}
which is identical to eq.~(\ref{Fishexact}).

\subsection{The Diagrammatic Mouse - where the UV/IR Mixing should occur}  \label{MouseFR} 

\begin{figure}[h]
\begin{center}
\begin{picture}(55,55)  
\put(5,5){\vector(0,1){40}}
\put(-5,45){\mbox{\small time}}
\put(48,20){\line(-4,1){25}}
\put(48,20){\vector(-4,1){15}}
\put(22,27){\line(1,1){7}}
\put(22,27){\vector(1,1){4}}
\put(32,34){\line(3,-2){18}}
\put(32,34){\vector(3,-2){9}}
\put(50,19){\mbox{$u$}}\put(51,20){\circle{6}}
\put(20,24){\mbox{$v$}}\put(21,25){\circle{6}}
\put(30,35){\mbox{$w$}}\put(31,36){\circle{6}}
\bezier{0}(19,23)(12,25)(19,27)\put(11,24){\mbox{$k_2^+$}}
\bezier{0}(29,38)(31,45)(33,38)\put(30,43){\mbox{$k_3^+$}}
\put(33,27){\mbox{$k_1$}}
\put(53,22){\line(1,1){10}}
\put(53,22){\vector(1,1){5}}
\put(53,18){\line(1,-1){10}}
\put(53,18){\vector(1,-1){5}}
\put(52,27){\mbox{$q_f^+$}}
\put(52,12){\mbox{$q_e^-$}}
\end{picture}
\end{center}
\caption{The macro-contribution $uvw$\label{mouse}}
\end{figure}
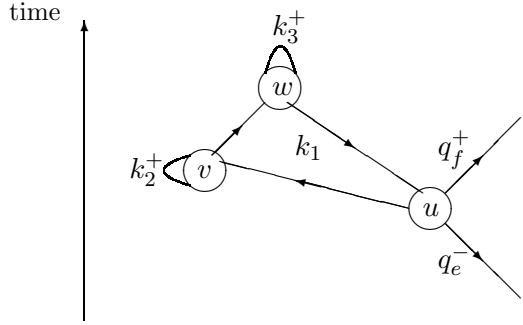

Confident in our new tools, we embark on calculating the two-point three-loop amplitude of ``mouse-like morphology'',
figure \ref{mouse}.

This amplitude is of great interest since in usual non-commutative QFT it is the simplest graph that becomes
undefined due to the notorious UV/IR-mixing problem: the two tadpoles inserted into the third each bring a $1/\tilde{k}_1^2$,
introducing a non-integrable IR-singularity into the remaining, otherwise UV-finite, loop-integral -- usually...

Beginning as in the previous sections (and skipping the steps that are now familiar), the amplitude of interest is written as
\begin{eqnarray}
&&\Gamma^{(2,3)} =\nonumber\\&&
\frac{i g^3}{(4!)^3} \int \frac{d^4k_1}{(2 \pi)^4} \frac{d^4k_2}{(2 \pi)^4} 
               \frac{d^4k_3}{(2 \pi)^4} \frac{1}{(2 \omega_1)^3} \frac{1}{2 \omega_2} \frac{1}{2 \omega_3}\nonumber \\
             & &  \sum_{\sigma_2,\sigma_3}^{+1,-1}  
                 \frac{1}{\sigma_2 k_2^0-\omega_2+i\varepsilon} \quad \frac{1}{\sigma_3 k_3^0-\omega_3+i\varepsilon}  \nonumber \\
           & & \sum_{\sigma_u,\sigma_v, \sigma_w}^{+1,-1}  
                     \Phi^v(k_2^+;k_1^{\sigma_u},  -k_1^{\sigma_v})  
                     \Phi^w(k_3^+;k_1^{\sigma_v},  -k_1^{\sigma_w}) 
                     \Psi^u(q_f^+,-q_f^+,  k_1^{\sigma_w},  -k_1^{\sigma_u}) \nonumber \\
           & & \frac{1}{\sigma_u k_1^0 - \omega_1 + i \varepsilon} \quad \frac{1}{\sigma_v k_1^0 - \omega_1 + i \varepsilon}
                    \quad \frac{1}{\sigma_w k_1^0 - \omega_1 + i \varepsilon} .
\end{eqnarray}
Two of the eight possible combinations of $\sigma_u=\pm,\sigma_v=\pm, \sigma_w=\pm$ 
result in the coincidence of all three poles on the same half of 
the complex plane, and they thus vanish under $k_1^0$-integration. The remaining six summands yield 
\begin{eqnarray}   \label{Mouse1}
&&\Gamma^{(2,3)}  =\nonumber\\&&
\frac{i g^3}{(4!)^3} \Big( -i \int \frac{d^3k_1}{(2 \pi)^3}\frac{1}{(2 \omega_1)^5} \Big)
                      \Big( -i \int \frac{d^3k_2}{(2 \pi)^3}\frac{1}{2 \omega_2} \Big) 
                      \Big( -i \int \frac{d^3k_3}{(2 \pi)^3}\frac{1}{2 \omega_3} \Big)\nonumber\\
 & & \Big(  \Phi^v(k_2^+;k_1^+, -k_1^+) \Phi^w(k_3^+, k_1^+, -k_1^-) \Psi^u(q_f^+,-q_f^+,k_1^-, -k_1^+) \nonumber\\
      && +  \Phi^v(k_2^+;k_1^+, -k_1^-) \Phi^w(k_3^+, k_1^-, -k_1^+) \Psi^u(q_f^+,-q_f^+,k_1^+, -k_1^+) \nonumber\\
      && +\Phi^v(k_2^+;k_1^-, -k_1^+) \Phi^w(k_3^+, k_1^+, -k_1^+) \Psi^u(q_f^+,-q_f^+,k_1^+, -k_1^-) \nonumber\\
      && +\Phi^v(k_2^+;k_1^+, -k_1^-) \Phi^w(k_3^+, k_1^-, -k_1^-) \Psi^u(q_f^+,-q_f^+,k_1^-, -k_1^+) \nonumber\\
      && + \Phi^v(k_2^+;k_1^-, -k_1^+) \Phi^w(k_3^+, k_1^+, -k_1^-) \Psi^u(q_f^+,-q_f^+,k_1^-, -k_1^-) \nonumber\\
      && +\Phi^v(k_2^+;k_1^-, -k_1^-) \Phi^w(k_3^+, k_1^-, -k_1^+) \Psi^u(q_f^+,-q_f^+,k_1^+, -k_1^-) \Big).
\end{eqnarray}
Here the six terms correspond to the six possible macro-time orderings of 
the vertices: $uvw, wuv, vwu, uwv, vuw, wvu$, respectively.

\section{No UV/IR Mixing in IPTOPT}  \label{Discussion} 

The most interesting feature of IPTOPT is the apparent absence of the UV/IR-mixing problem.
This can be seen in the amplitudes calculated so far by explicitly performing the loop integrations in the 
result of the previous section \ref{MouseFR}. 
No divergence will be fed down via the phases to the next loop.

\subsection{Explicit Result for the UV/IR Divergence-free Mouse}

To evaluate eq.~(\ref{Mouse1}) explicitly, we start by integrating over $k_2$ and $k_3$.
These integrals yield, apart from a possible overall cosine in $k_1$, exactly the $\mathcal{I}$'s 
from section \ref{NcFullProp} and \cite{Noncom1}:
\begin{eqnarray} 
& & \int \frac{d^3k}{(2 \pi)^3 }\frac{1}{2 \omega} \Phi(k^+;k_1^{\sigma}, 
-k_1^{\sigma'})  \nonumber \\
&=& \int \frac{d^3k}{(2 \pi)^3 }\frac{2}{2 \omega} \cos\Big(\frac12k_1^{\sigma} \tilde  k_1^{\sigma'}\Big) 
             \Big( 3 + e^{-i k^+ \tilde{k}_1^{\sigma}} + e^{i k^+ \tilde{k}_1^{\sigma'}} + 
                    e^{-i k^+ (\tilde k_1^{\sigma}-\tilde k_1^{\sigma'})} \Big) \nonumber \\
&=& 2 \cos\Big(\frac12 k_1^{\sigma} \tilde  k_1^{\sigma'}\Big)\mathcal{I}(k_1^{\sigma},-k_1^{\sigma'}).
\end{eqnarray}
Since these were already evaluated in eq.~(\ref{I++})--(\ref{Ineu}), determining the result for all but the last 
loop integration is a mere task of compilation. Using the same abbreviations as above  ($k_1 \rightarrow k$) we find
\begin{eqnarray} \label{MouseFR3} 
&&\Gamma^{(2,3)} =\nonumber\\&& -\frac{g^3}{4 (3!)^3} \frac{1}{(2 \pi)^4} \int \frac{d^3k}{(2 \pi)^3 }\frac{1}{(2\omega)^5} 
           \Big[ 1 + \cos(k^+ \tilde{k}^-) \Big] \nonumber\\
&& \times\Big[ \frac32\mathcal{Q} - \frac{2m}{|\tilde{k}^+|} K_1(m|\tilde{k}^+|)- \frac{2m}{|\tilde{k}^-|} K_1(m|\tilde{k}^-|) 
                + \frac{m}{\omega_k |\Theta_{0i}|} K_1(2m \omega_k |\Theta_{0i}|)    \Big]  \nonumber \\
&& \times  \Big[ \Big( 4 + \cos(k^+\tilde{q}_f^+) + \cos(k^-\tilde{q}_f^+) \Big) \nonumber \\
& &\ \times \Big( \frac32\mathcal{Q} - \frac{2m}{|\tilde{k}^+|}K_1(m|\tilde{k}^+|)- \frac{2m}{|\tilde{k}^-|}K_1(m|\tilde{k}^-|) 
                   + \frac{m}{\omega_k |\Theta_{0i}|} K_1(2m \omega_k |\Theta_{0i}|)   \Big) \nonumber \\
& &\ +\quad\Big( 3 + \cos(k^+\tilde{q}_f^+) + \cos(k^-\tilde{q}_f^+) + \cos((k^+ - k^-)\tilde{q}_f^+) \Big) \nonumber \\
& &\  \times\Big( 2 \mathcal{Q} - \frac{m}{|\tilde{k}^+|} K_1(m|\tilde{k}^+|)- \frac{m}{|\tilde{k}^-|}K_1(m|\tilde{k}^-|)\Big) 
\Big]
\end{eqnarray}
where $K_1(x)$ is the modified Bessel function. 
So far in non-commutative QFT this expression contained an IR divergence: poles in $k^2$ of $2^{\textrm{nd}}$ 
order. 
This is not the case here as
\begin{equation} \label{IRlimit}
 \lim_{\vec{k} \to \vec{0}} (\tilde{k}_{\mu}^{\pm})^2 = - m^2 \Theta_{0i}^2 <0.
\end{equation}
This limit will be discussed in more detail in the next section.

Where does this first instance of the absence of the notorious UV/IR-mixing problem stem from?
It is due to the appearance of on-shell 4-momenta in the non-commutative phases: since, because of
the mass, the 0-component remains non-vanishing for all values of the three-momentum, no pole can appear.

\subsection{Argument for the General Absence of UV/IR Mixing} \label{NoUVIR}

Encouraged by the above explicit result we give an argument for the absence of this problem to all orders 
--- for all $\Gamma^{(n,l)}$ --- in (IPTO) perturbation theory in a more general way 
(although we refrain from writing ``proof'').

To arrive at this conclusion we remember the $n$-th order $k$-point Green functions 
given by the Gell-Mann--Low formula 
\begin{eqnarray} \label{proof1}
 &&G_n(x_1,\ldots, x_k) \\
&&\qquad = \frac{i^n}{n!}\int d^4z_1 \ldots d^4z_n 
           \langle 0 |  T \phi(x_1)\ldots \phi(x_k) \mathcal{L}_I(z_1)\ldots \mathcal{L}_I(z_n)  | 0 \rangle^{con},\nonumber
\end{eqnarray}    
where $T$ denotes the time ordering  and $\mathcal{L}_I(z)$ is the interaction part of the Lagrangian, 
$\frac g{4!} (\phi \star \phi \star \phi \star \phi)$ for non-commutative $\phi^4$-theory.

Note that all fields occurring in eq.~(\ref{proof1}) are \emph{free fields},
their Fourier transforms are on-shell quantities, 
 the 0-component of the four-vector being $\omega(\vec{k})$:
\begin{equation}  \label{freefield}
\phi(x)=\int \frac{d^3k}{(2\pi)^3} \Big( \tilde{\phi}(\vec{k}) e^{-i k^+ x} + \tilde{\phi}^{\dagger}(\vec{k}) e^{+i k^+ x} \Big).
\end{equation}
Evaluating the $\star$-product between these FT free fields hence produces phase factors containing \emph{on-shell}
momenta $k^\pm_{\mu}$ only (see also the discussion in chapter 3 of \cite{Noncom1}). 
This remains true after integrating out some (or all) of the loop-momenta occurring later in the
evaluation. At no point of the further calculations (evaluating TO, FT, amputation, ...) will this property be changed.  

Why does this novel feature of IPTOPT prohibit the occurrence of the usual UV/IR problem? 
First note that for timelike (on-shell) four vectors $k^{\mu}$ we find  $\tilde{k}^{\mu}$ to be spacelike
\begin{equation} \label{proof2}
\Theta_{\mu \nu} k^{\mu} k^{\nu} = 0 =  (k^{\mu} (\Theta_{\mu \nu}) k^{\nu} := \tilde k^{\mu} {k}_{\mu}
\end{equation}
and hence
\begin{equation} \label{proof3}
\tilde{k}^{\mu} \tilde{k}_{\mu} < 0 \quad \forall \tilde{k}^{\mu} \neq 0^{\mu} \quad, \quad 
\tilde{k}^{\mu} \tilde{k}_{\mu} = 0  \leftrightarrow \tilde{k}^{\mu} = 0^{\mu}.
\end{equation}
The case $\tilde{k}^{\mu} = 0^{\mu}$ is only possible for massive theories iff $\Theta_{\mu \nu}$ is of less then full rank, 
which is excluded in IPTOPT since we demand $\Theta_{i0} \neq 0$: if $\Theta_{\mu \nu}$ were of less than full rank, one 
could always transform it into $\Theta'_{\mu \nu}$, with $\Theta'_{i0} = 0$, which we excluded by definition.

Hence we find
\begin{equation} \label{proof4}
\tilde{k}^2 = \tilde{k}^{\mu} \tilde{k}_{\mu} < 0   \quad \forall \vec{k}.
\end{equation}

As the usual (i.e. the one found in the literature)  UV/IR problem always occurs in the form of a $1/\tilde{k}^2$ pole,
which, for off-shell $k^{\mu}$ and $\tilde{k}^{\mu}$, introduces a possible new singularity at 0, we see that 
IPTOPT is free from this (type of) problem: zero is never reached by  $\tilde{k}^2$. 

It is at this point that our argument degrades from being a proof, since it excludes the appearance of this particular
form of mixing only. But in what other guises it still has to be excluded we are not able to discuss yet.

\subsection{A short Note on $PT$} \label{PT}

As a short side-remark we would like to draw your attention to the behaviour of the amplitudes calculated above under
$P$ and/or $T$ acting on the external momenta:
\begin{equation}
P: \quad \vec{q} \to -\vec{q}, \qquad T: \sigma \to -\sigma.
\end{equation}
Hence we find that
\begin{equation}
P(q^{\pm}) = -q^{\mp}, \qquad T(q^{\pm}) =q^{\mp}, 
\end{equation}
which do not leave the amplitudes calculated above invariant when only one of $P, T$ acts on them.
However, under the combined action of $PT$: 
\begin{equation}
PT(q^{\pm}) = P(q^{\mp}) = -q^{\pm}, 
\end{equation}
the amplitudes remain unchanged, since the external momenta occur in cosine only.

Invariance under $PT$, however, is a direct consequence of the unitarity of the S-matrix and the existence of 
free states; see \cite{Alvarez-Gaume03} and references therein.

\section{Outlook}   \label{Outlook}

In this article one further step was taken in the program of IPTOPT: 
Feynman rules were stated and demonstrated to yield the same results as the TO amplitude.
In a sense, IPTOPT developed into interaction-point diagrammatics.
Although note must be taken that these FR are rather conjectured than truly derived, 
since (Minkowskian) canonical instead of (Euclidean) PI quantization was employed.
A more general and rigorous method for obtaining them has recently been found in \cite{Denk}. 

Also a strong motivation for further work utilizing this approach was discovered:
the possibility of the  general absence of the UV/IR problem.
Although a strong argument in favour of this feature was given, a true proof is still missing and certainly highly desirable. 
In principle two routes to this end are imaginable: either continuing in IPTOPT, investigating eq.~(\ref{GammaE})
for the possibility of an inductive proof;
or by making use of the diagrammatics proposed in this work.
The second approach could also yield important insights into how to pursue the great question of renormalizability and
renormalization of non-commutative QFT.  

Further work may deepen our understanding of the intricate connections between nano-causality, 
unitarity, UV/IR mixing (i.e. its absence), CPT invariance and renormalization. Moreover,
possible phenomenological implications of IPTOPT will be of great interest \cite{Liao:2002kd}. Anyway,
with non-commutative QFT 
a tool to a better understanding of commutative QFT is available, illustrating by similarities and differences
the fundamental features of the two sets of theories.

\subsection*{Acknowledgements}

VP would like to thank Prof.~M.~Schweda and R.~Wulkenhaar for helpful 
discussion and support. 
PF would like to thank C.~Jarlskog, T.~Hurth,  L.~\'Alvarez-Gaum\'e and A.~Bichl for lectorship 
and discussions, S.~Vascotto for excellent and thorough linguistic corrections.

\end{document}